\documentclass[journal]{IEEEtran}
\usepackage{amsmath,amssymb,amsfonts}
\usepackage[utf8]{inputenc}
\usepackage{graphicx}
\usepackage{todonotes}
\usepackage[bookmarks=false]{hyperref}
\usepackage[author={Max Schlepzig}]{pdfcomment}
\hypersetup{colorlinks=false,pdfborder={0 0 0},hypertexnames=false}
\usepackage{algorithm}
\usepackage[noend]{algpseudocode}
\usepackage{subfig}
\usepackage{textgreek}
\usepackage{gensymb}
\usepackage[font=small]{caption}

\newcommand{\orcid}[1]{\href{https://orcid.org/#1}{\includegraphics*[width=10pt]{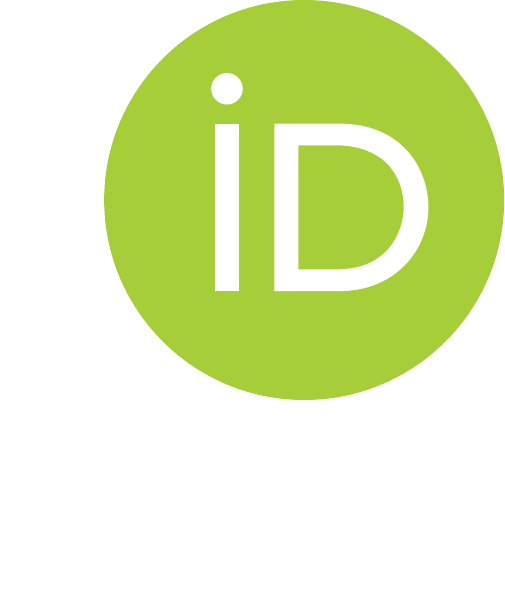}}}

\begin{document}
%
\title{An Error-Based Approximation Sensing Circuit for Event-Triggered Low-Power Wearable Sensors}
%
%
%

\author{\IEEEauthorblockN{%
    Silvio~Zanoli\orcid{0000-0002-0316-1657},\hspace{2mm}
	Flavio~Ponzina\orcid{0000-0002-9662-498X},~\IEEEmembership{Student member,~IEEE}\hspace{2mm}
	Tomás~Teijeiro\orcid{0000-0002-2175-7382},\hspace{2mm}
	Alexandre~Levisse\orcid{0000-0002-8984-9793},~\IEEEmembership{Member,~IEEE}\hspace{2mm}
	and~David~Atienza\orcid{0000-0001-9536-4947},~\IEEEmembership{Fellow,~IEEE}%
}

\IEEEauthorblockA{Embedded Systems Laboratory (ESL), EPFL, Lausanne, Switzerland}%
\thanks{Corresponding author: S. Zanoli (email: silvio.zanoli@epfl.ch).}}

%
%

\markboth{IEEE}%
{Ponzina \MakeLowercase{\textit{et al.}}: An error-based approximation ADC circuit for event-triggered, low power wearable sensors}
%



\maketitle

\begin{abstract}
Event-based sensors have the potential to optimize energy consumption at every stage in the signal processing pipeline, including data acquisition, transmission, processing, and storage. However, almost all state-of-the-art systems are  still built upon the classical Nyquist-based periodic signal acquisition. In this work, we design and validate the Polygonal Approximation Sampler (PAS), a novel circuit to implement a general-purpose event-based sampler using a polygonal approximation algorithm as the underlying sampling trigger. The circuit can be dynamically reconfigured to produce either a coarse or detailed reconstruction of the analog input by adjusting the error threshold of the approximation.
The proposed circuit is designed at the Register Transfer Level and processes each input sample received from the analog-to-digital converter (ADC) in a single clock cycle.
The PAS has been tested with three different types of archetypal signals captured by wearable devices (electrocardiogram, accelerometer, and respiration data) and compared with a standard periodic ADC. These tests show that single-channel signals, with slow variations and constant segments (like the used single-lead ECG and the respiration signals), take great advantage of the used sampling technique, reducing the amount of data used up to 99\% without significant performance degradation. At the same time, multi-channel signals (like the six-dimensional accelerometer signal) can still benefit from the designed circuit, achieving a reduction factor of up to 80\% with minor performance degradation. These results open the door to new types of wearable sensors with reduced size and higher battery lifetime.
\end{abstract}


%
\IEEEpeerreviewmaketitle

\section{Introduction}
%
%
%
%

\IEEEPARstart{S}{mart} sensor data and edge-computing internet of thing (IoT) devices are the building blocks for the new paradigm that is Industry 4.0 and its future. Interestingly, this new way of conceptualizing work has the potential to effectively fuse together industry and healthcare in what is often referred to as Healthcare 4.0~\cite{industry_and_healthcare}. This term will enable a much healthier society as well as more productive industries. 

In this context of Industry 4.0, wearable devices have the potential to optimize human workflow~\cite{industry_survey,industry_tools_user_exp} while monitoring and securing the health of every worker, creating a win-win situation for both the industry sector and the workforce. However, behind this goal of effective continuous monitoring, the designers of wearable devices are faced with a permanent struggle to optimize energy consumption and extend battery lifetime. Also, as edge computing gains momentum, wearables are expected to go beyond the typical sensing and transmitting functionalities to more advanced in-sensor analysis~\cite{Varghese2016}; thus, adding an additional energy consumption factor. 

The battle for energy reduction has been waged on various fronts. In particular, different works deal with data storage, low-power data processing architectures, and low-energy wireless communication~\cite{Wei19, Abadal20, Panades20, Pullini18,low_pow_offloading}. However, data sensing, even if remains the largest energy expenditure factor in many applications~\cite{rincon2011development}, has not undergone such a significant evolution. Event-based sampling has proven large potential energy savings in a few wearable-based applications~\cite{rovere20182,surrel20}, but no framework has been proposed to make it applicable in a general way to existing sensing systems in the IoT context. 
The lack of applications that use event-based sampling strategies is due to the fact that the research community is just now starting to explore the capabilities and limitations of this field.

In this work, we propose a novel hardware module, named Polygonal Approximation Sampler (PAS), that can be associated with a regular analog-to-digital converter (ADC) in order to migrate a regular sensor to an event-triggered low-power strategy. Conceptually, this module acts as a sample selector, taking as input the regularly acquired samples from the ADC, and outputting the minimum set of samples required to reconstruct the signal within a locally bounded error. The full technique is discussed in Section~\ref{sec:algorithm}, with additional context on event-based signal processing and our contribution to adapting Finite Impulse Response (FIR) filters to this domain.

In recent years, we have observed the rise of new event-based sampling techniques, which can be grouped into three areas of work: 1) variations of the level-crossing sampling method \cite{rovere20182,Level_crossing_comparison,Tang2013}, 2) continuous time level-crossing analysis \cite{CT_level_crossing}, or 3) sampling optimization based on compressed sensing \cite{Cmpressed_sensing_matrix}, and subsequent optimization of the used measurement matrix \cite{CS_dynamic_knob}. Our approach generalizes the event-based acquisition problem, defining an objective function (in our case, the maximum allowed integral error) and obtaining the relevant events in an adaptive manner. This is opposed to the rigid sampling framework imposed by level-crossing, while, at the same time, it remains signal-agnostic, contrary to compressed sensing approaches.

Our proposed design and hardware implementation of the PAS, as well as its logical interconnections with the ADC and the processor, are described in detail in Section~\ref{sec:wds}. Then, in Sections~\ref{sec:applications} and~\ref{sec:results}, we explore the potential benefits of the proposed circuit from an application perspective. In the wearable domain, we are targeting energy-hungry applications aimed at continuous monitoring of biomedical signals. These applications usually work in a windowed fashion over long streams of data, and memory is the primary responsible for energy expenditure~\cite{Zanoli20}. The objective of this experimental validation is to characterize the trade-off between the amount of input data that can be removed using our event-based approach, and the impact this has on the performance of the target task. To this end, we have selected three different types of tasks and signals that are typically targeted with wearable devices:
the detection of QRS complex (i.e., the Q, R, and S waves), which represents one of the main features of the electrocardiogram (ECG) signals
, task recognition in inertial measurements, and breath estimation from respiratory signals. In this way, we demonstrate that the PAS has general applicability in multiple scenarios and with signals of different nature coming from different devices. Moreover, as shown in~\cite{industry_activity_recon,industry_evaluation_through_sensors,industry_tools_user_exp}, these tasks have become more and more important for workflow organization in the last few years, making them prime candidates for smart-sensing optimization.

Finally, in Section~\ref{sec:system_integration}, we explore a complete system-level integration of the PAS with a low-power Micro-controller unit (MCU), and we experimentally assess the potential energy savings that can be achieved in a real application. For this, we take the QRS complex detection task as our benchmark use case.

We can hence summarize the main outcomes of our work as follows:
\begin{itemize}
    \item Design of a polygonal approximation sampler (PAS).
    \item Characterization of the PAS hardware implementation.
    \item Study of the effect of event-based(EB) sampling using PAS on the long-term analysis of various biosignal.
    \item System integration test for energy/performance characterization of the proposed PAS implementation.
\end{itemize}

\section{Background on Event-Based Sampling}
\label{sec:algorithm}

In this section, we describe a new approach to event-based sampling, and the implemented method grounded on the work of Wall and Danielsson~\cite{Wall1984}, in which a curve is polygonally approximated based on a control variable. Then, we briefly describe how this new environment conditions the use of classical signal processing techniques, with a focus on filtering operations.

\subsection{Polygonal approximation and the Wall-Danielsson method}
\begin{figure}[ht]
	\centering
    \includegraphics[width=1.0\linewidth]{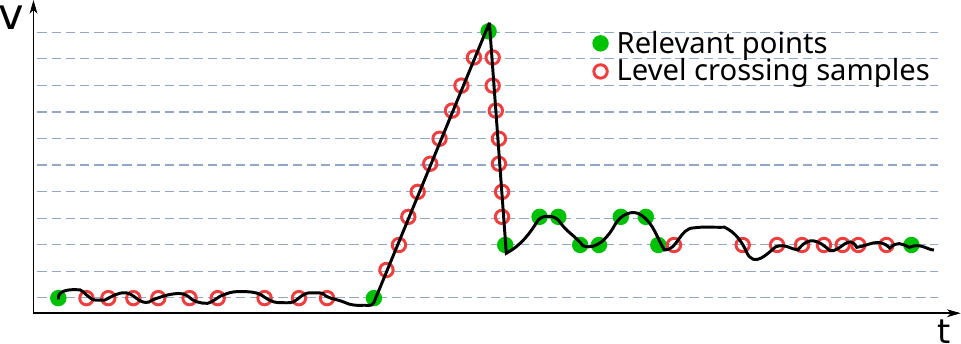}
	\caption{Limitations of the level-crossing approach for event-based sampling. \small{[Blue dashed lines represent sampling levels, green dots are the actual relevant samples, and red circles correspond to unnecessary samples that are avoided with polygonal approximation]}}
	\label{fig:level-crossing}
\end{figure}

The usual approach to event-based sampling for bio-signal monitoring is the level-crossing method~\cite{Tang2013}. The idea is to sample just when the signal crosses certain amplitude values, that can be more or less densely separated depending on the target resolution. This scheme is efficient and easy to implement, but it has two main limitations, as illustrated in Fig.~\ref{fig:level-crossing}. First, if the signal oscillates near one of the levels, it is sampled much more frequently than if it oscillates between two levels. Second, if the signal has high-amplitude linear variations, this approach results in a number of samples equivalent to the number of levels crossed by each variation, whereas only two samples would be sufficient to represent each of the variations. 
Moreover, while several adaptive compressed sensing techniques have been developed in the previous years, showing excellent results \cite{CS_dynamic_knob}, such methods rely on strong assumptions on the signal under analysis and require a different pre-processing pipeline (and hence hardware) for every type of signal under analysis.
Therefore, we propose an approach for low-power event-based sampling on wearable sensors that is aimed at representing the input signal with the minimum number of points, assuming a linear behavior between each pair of consecutive points.

From the plethora of available algorithms~\cite{Keogh2004}, we inspired our circuit design on the Wall-Danielsson method~\cite{Wall1984}, as it is particularly suitable for a real-time hardware implementation. This algorithm works by ensuring that the discrete integral distance between the uniformly sampled signal and the linear interpolation between two event-based samples never grows more than a defined threshold $\epsilon$. On the one hand, it has the following interesting properties:

\begin{enumerate}
 \item It has linear complexity.
 \item The memory footprint is extremely small, requiring to store merely five numeric variables.
 \item It only requires one sample of the original signal in advance to estimate the approximation error.
\end{enumerate}

On the other hand, it does not guarantee optimality in the number of generated samples, but as we show in Section~\ref{sec:results}, it achieves a very high sampling reduction factor (i.e., the percentage of processed input elements with respect to the total number of input samples) in real application scenarios.
Note that, in this context, the aforementioned sampling reduction factor is not directly related to the relaxation of the sampling frequency of the input signal. Instead, it is the result of the proposed hardware implementation of the Wall-Danielsson algorithm, where the sampled input signal is \emph{pre-processed}, ultimately making the number of samples reaching the $\mu P$ lower than the total number of input samples. A more detailed description of this solution is presented in Section~\ref{sec:wds}.  

\subsection{Adapted Signal Processing Routines}
\label{subsec:methods}
By using the Wall-Danielsson algorithm to sample a signal, we lose the main assumption of classical signal processing: the uniformity of time-based sampling. Thus, in order to exploit the advantages given by the non-uniformity of the sampling, we need to adapt several of the most common operations needed for signal processing, namely FIR filtering, Fourier analysis, and statistical signal processing operations. In the following, we provide an illustrative example of how FIR filtering is performed, so it can help to understand how the target experiments were replicated.

\begin{figure}[tp]
  \centering
  \includegraphics[width=0.9\linewidth]{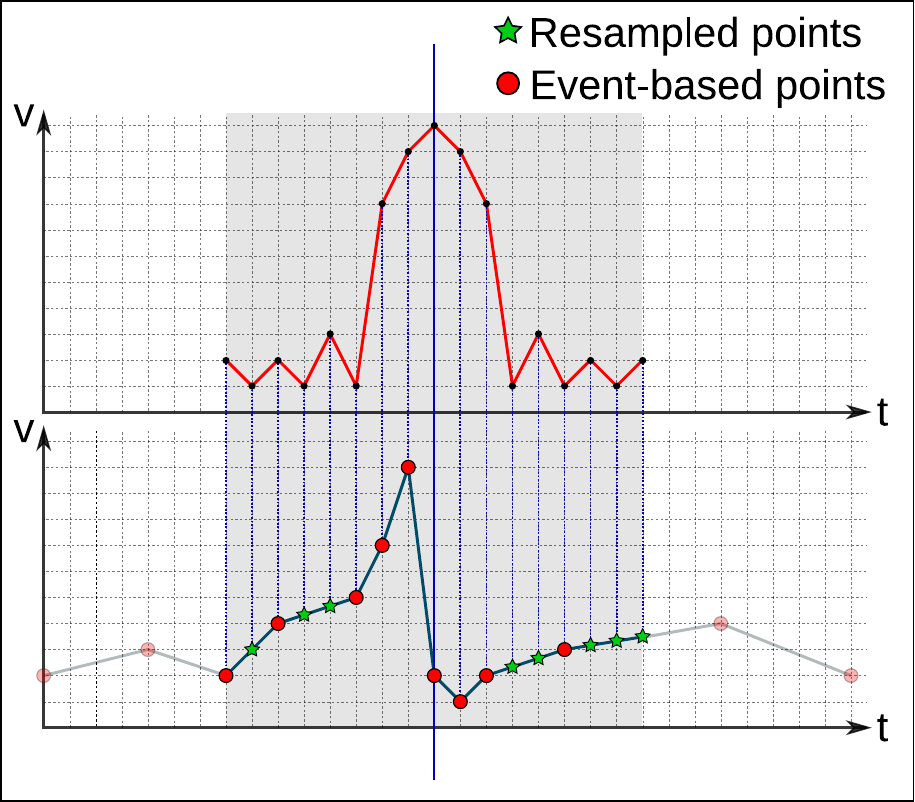}
  \caption{
How a FIR filter (coefficients in the top graph) is applied to an event-based signal (low graph). This figure highlights the limited number of points that need to be interpolated (stars).
}
  \label{fig:FIR}
\end{figure}

One of the most common techniques in signal processing is the filtering operation. Fig. \ref{fig:FIR} shows the idea behind the implementation of FIR filtering we performed: the FIR filters are only applied to the event-based samples (from here on, referred just as samples) and linearly interpolate only the points strictly needed for the application of the filter on that point. By doing this, we only need to compute the FIR filter for a fraction of points defined by $\frac{P_{sampled}}{P_{all}}$, where $P_{sampled}$ is the number of points sampled with our proposed technique, and $P_{all}$ is the number of points of the signal acquired using the classical uniform sampling rate. Note that, whenever we apply the filter to a point, we need to interpolate, at most, the number of coefficients in the FIR filter minus one.

The adapted signal processing routines work relative to each sample or relative to a defined time span. Hence, changes in the average sampling frequency (i.e., changes in the threshold of the Wall-Danielsson algorithm) do not affect the average amount of operations per sample to be executed.

\section{Polygonal Approximation Sampler System Design}
\label{sec:wds}

\subsection{Overview}
Here, we introduce our proposed solution to improve signal processing efficiency on a plethora of wearable-based applications. We designed a hardware module, named PAS, that acts as a general-purpose event-based sampler performing a polygonal approximation computation.
It interacts with both the ADC and the microprocessor/microcontroller ($\mu$P), and it aims at lowering the workload of the $\mu$P by filtering the samples received from the ADC, thus decreasing the amount of data to be processed. This approach opens the path for a variety of energy savings techniques. On one side, the reduced workload allows the $\mu$P to spend more time in an energy-efficient sleep mode. 

On the other side, the reduction of acquired samples that need to be processed in each window enables the use of smaller memory blocks, ultimately reducing leakage energy. Indeed, the lower number of samples that have to be processed by $\mu P$ in a certain input window (for example, we consider 20-second windows in our experiments in Section~\ref{sec:system_integration}) require smaller memory buffers for their storage than the total number of acquired inputs.

In this section, we first focus our attention on the architectural structure and computational behavior of the proposed module. Then, after synthesizing it, we present its energy, area, and performance evaluation. Finally, a complete overview of the energy gains that can be achieved by introducing this module is presented in Section~\ref{sec:system_integration}, where we evaluate our benchmarks on a fully-integrated system, including both the PAS and a low-power $\mu$P.

\begin{figure}[tp]
	\centering
    \includegraphics[width=1.0\linewidth]{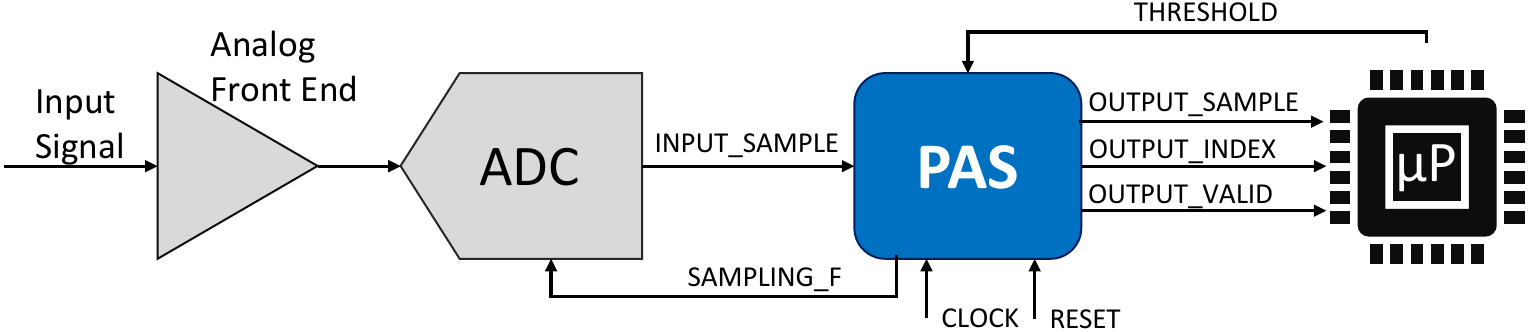}
	\caption{Block diagram of the overall architecture. The Polygonal Approximation Sampler (PAS) receives the ECG samples from the ADC and sends the computed approximated signal to the $\mu$P. }
	\label{fig:architecture}
\end{figure}

\subsection{System-Level Block Diagram}
As shown in Fig.~\ref{fig:architecture}, the PAS lies in between the ADC and the processor: it captures the samples from the ADC and broadcasts to the processor those that are selected according to the polygonal approximation algorithm. The threshold parameter ($\epsilon$), presented in Algorithm~\ref{alg:wds_algo}, is received from the processor and allows the PAS to adapt the algorithm sensitivity, thus enabling a dynamic optimization of the sampling rate at the application level. 
Nonetheless, the selection of the threshold $\epsilon$ is signal and application dependent and, therefore, it must be determined for each specific problem. However, as a general rule of thumb, this threshold should be approximately equal to the square of the minimum amplitude variation producing a significant change in the input signal.

Using this architectural solution, we can filter the samples from the ADC, and forward to the processor only the ones selected by the algorithm. The result is a reduction of the processor computational load (more than 60\%, as discussed in the next sections), leading to a reduction in the overall system energy consumption, as the processor can be put in a low-power state when no samples need to be processed.


Fig. \ref{fig:architecture} also shows the input/output signals of the PAS. \textit{OUTPUT\_SAMPLE} and \textit{OUTPUT\_INDEX} characterize the input samples that are forwarded to the processor. In particular, \textit{OUTPUT\_SAMPLE} holds the signal value received from the ADC, while \textit{OUTPUT\_INDEX} represents the index difference between two consecutive forwarded samples. Additionally, \textit{OUTPUT\_VALID} is used as a wake-up signal for the processor, so that it can activate the corresponding interrupt service routine (ISR) to read the input sample only when needed, remaining idle the rest of the time. While these three output signals are connected to the processor, \textit{SAMPLING\_F} is sent to the ADC to drive its working frequency. We assume a bitwidth of 16 bits for the received samples, which allows the input signal to be correctly represented in our benchmarks (i.e., no saturated values appear).

\subsection{PAS functional behavior}
\label{sec:pas_behavior}

\begin{algorithm}[tp]
\footnotesize
\caption{PAS Procedure. $x$ and $y$ represent the cumulative $xy$ coordinates of the acquired input samples, with $dx$ and $dy$ being the increment/decrement of the corresponding coordinates of the current sample with respect to the previous one, respectively.  PAS forwards the acquired input sample only when the cost function $f$ exceeds the imposed threshold $\epsilon$, providing $\mu P$ with the sample value and the time-stamp difference with respect to the previous output (i.e., $\overline{t}-t$).
}

\label{alg:wds_algo}
\begin{algorithmic}[1]
\Procedure{PAS}{threshold $\epsilon$}
    \State $dx=1$
    \State $i=1$
    \State $f=x=y=\overline{t}=peak=length=0$
    
    \Loop
        \State $dy=sample[i]-sample[i-1]$
        \State $x=x+dx$
        \State $y=y+dy$
        \State $f=f+x*dx-y*dy$
        \State $displacement=abs(y)+x$
        
        \If {$displacement<length$ and $peak=0$}
            \State $peak=peak+i-1$
        \EndIf
        
        \State $length=displacement$
        
        \If {$abs(f)>\epsilon$}
            \If {$peak=0$}
                \State $t=i-1$
            \Else
                \State $t=peak$
            \EndIf
            
            \State $output(sample[t], t-\overline{t})$
            \State $f=peak=0$
            \State $x=(i-t)*dx$
            \State $y=sample[i]-sample[t]$
            \State $\overline{t} = t$
            \State $length=abs(y)+x$
        \EndIf
        \State $i=i+1$

    \EndLoop
\EndProcedure
\end{algorithmic}
\end{algorithm}

\begin{figure}[tp]
    \includegraphics[width=1.0\linewidth]{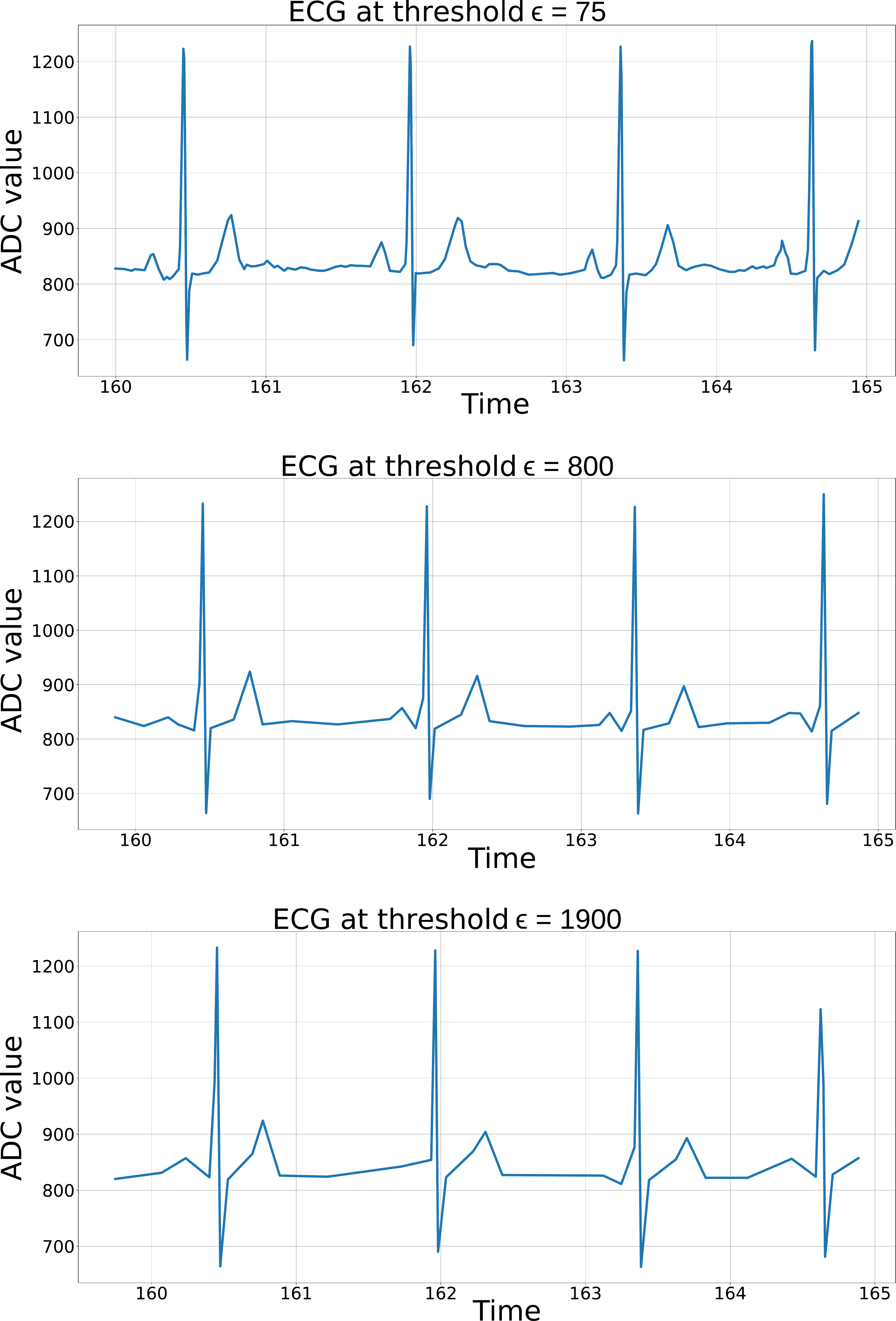} 
    \caption{
ECG signal, sub-sampled with the proposed algorithm, using three different $\epsilon$ threshold values.
}
    \label{fig:subTh}
\end{figure}

Algorithm~\ref{alg:wds_algo} shows the pseudocode of the adapted Wall-Danielsson algorithm, as implemented in our hardware circuit design. The $\epsilon$ parameter allows us to control the coarseness of the approximation, as illustrated in Fig.~\ref{fig:subTh}. The possibility of tuning this parameter in real-time allows us to regulate the sampling toward a more coarse or fine representation of the signal, depending on the local properties of the signal itself.
Although the sampling frequency is dictated by the PAS and can be dynamically configured at run-time, we consider it fixed for most of the time. This assumption enables an algorithmic simplification since the horizontal step $dx$ becomes constant and can be set to 1 (see line 2 of Algorithm~\ref{alg:wds_algo}). In this way, arithmetic instructions involving this variable reduce their complexity, as additions simplify to unitary increments and multiplications can be avoided, resulting in a more energy efficient and faster HW implementation.

Finally, the use of the index difference between two consecutive forwarded samples instead of the absolute sample index (that is, \textit{OUTPUT\_INDEX} in Fig. 3) has two main advantages. First, it enables the recording of a potentially infinite sequence of input samples. Secondly, it still permits the tracking of time references as it indeed counts the input samples received from the ADC. To prevent numerical overflow, we reset the counter when it reaches the highest representable value and send the corresponding input sample. This condition may arise when the input signal is constant or, more generally, shows fluctuations lower than the imposed threshold $\epsilon$.

\begin{table}[tp]
\caption{Area and power details of synthesized PAS.}
\label{tab:synthesis_info}
\centering
\begin{tabular}{lc}

\hline\hline
\multicolumn{2}{c}{Area report} \\
\hline\hline
Combinational cells & $1220$ \\
Non-Combinational cells & $144$ \\
Combinational Area &  $ 2083 nm^2$ \\
Non-Combinational Area & $ 508 nm^2$ \\
\hline
 & \\
\hline\hline
\multicolumn{2}{c}{Power report} \\
 \hline\hline
Dynamic Power & $1.242\mu W$ \\
Leakage Power & $0.428 \mu W$ \\
Comb. logic power & $83.9\%$\\
Non-Comb. logic power &  $16.1\%$\\
\hline

\end{tabular}
\end{table}

\subsection{RTL Implementation}
The PAS is implemented at the Register Transfer Level (RTL) in VHDL. We define five combinational processes that allow the PAS to execute its functions in one clock cycle. 
Although the polygonal approximation algorithm is executed in several steps and iterations, it mainly consists of simple arithmetic instructions that do not make it a compute-intensive procedure. 
In fact, post-synthesis evaluations reveal that, even relaxing the timing constraints, the output signals become stable in less than $10~ns$ after the rising edge of the clock, meaning that the proposed PAS can also work also with higher-frequency signals up to $100~KHz$, not analyzed in this work.
 
We evaluate the correctness of our synthesized PAS with cycle-accurate RTL and post-synthesis simulations. In particular, the sampler is fed with input signal segments from the proposed benchmarks. The output signals are then compared with the results of a python-based implementation of the same algorithm, to validate the correctness of the designed circuit. Simulations are run using Mentor Graphics ModelSim\footnote{https://eda.sw.siemens.com/en-US/ic/modelsim}, while synthesis is run using Synopsys Design Compiler\footnote{https://www.synopsys.com/implementation-and-signoff/rtl-synthesis-test/design-compiler-graphical.html}, using the TSMC $40~nm$ technology library. 

Synthesis details are summarized in Table \ref{tab:synthesis_info}.
The synthesized sampler exhibits an area footprint of $0.0025~\mu m^2$, mainly composed of combinational cells. We also observe that 74.2\% of the area footprint implements arithmetical operators, such as adders, and the multiplier involved in the computation of the current segment length and its comparison with the imposed threshold. 
The estimated power consumption of $1.6 \mu W$ is dominated by the switching activity of combinational cells ($>80\%$), although the synthesis process that maximizes leakage is selected. More details about the energy contribution of our synthesized circuit are presented in Section~\ref{sec:system_integration}, where we illustrate a full system simulation. 

\section{Target applications}
\label{sec:applications}

In order to assess the applicability of our proposed PAS design to real-world signals, we propose an experimental setup in which we analyze the potential benefits of the described sampling technique on three different types of bio-signals: Electrocardiography (ECG), Inertial measurements, and Impedance Respiratory (IR) signals. For each of these signals, we consider a different target task to be performed: QRS complex detection from the ECG, task recognition from the inertial measurements, and breath rate detection from the IR signal.

For the analysis, we consider a state-of-the-art algorithm for each of the tasks, and we assess the performance penalty observed in our proposed event-based framework, contrasting it with the benefits in terms of reduced data volume. The key metric in this sense is the sampling reduction factor (SRF), defined as 
\begin{equation}
SRF = 1- \frac{Avg.~Event\_Based~freq.}{Uniform~freq.}
\end{equation}

where 
\begin{equation}
Avg.~Event\_Based~freq = \frac{Number~of~Events}{Signal~time~duration}
\end{equation}
This measure tells us, on average, the percentage of samples pruned from the uniformly sampled signal.

All the experiments were performed analyzing linearly increasing values of the threshold $\varepsilon$, since $SRF$ is largely dependent on its value.

In the remaining of this section, we explain the specific event-based implementations of the algorithms for QRS detection in ECG signals, task recognition from inertial signals, and breath rate estimation from IR signals. First, we give an overview of the used dataset. Then we describe the algorithm used to process the uniform-sampled signals and the methods used to port them to the event-based domain. The results of these experiments are presented in Section~\ref{sec:results}.

\subsection{QRS detection in ECG signals}
\label{subsec:QRS_ECG}
The QRS complex is one of the most prominent features of the ECG signal, as it is the signature of the ventricular contraction in the heart. Hence, QRS complex detection is the first step for heart rate analysis. The experiment here described is the natural extension of the work performed in \cite{Zanoli:279647} where the objective was to implement a QRS detection algorithm for event-based ECG signals. In this prior work, we analyzed the impact of level crossing sampling \cite{rovere20182} on energy and $F_1$ score performances.
Where the $F_1$ score is defined as the harmonic mean between precision and recall, eq.~\ref{eq:F_1-def} express it in terms of true positive($TP$), false positive($FP$), and false negative($FN$).
\begin{equation}
    F_1 = \frac{2TP}{2TP+FP+FN}
    \label{eq:F_1-def}
\end{equation}
In the context of QRS detection, we consider true positive as the correct identification of the complex, false positive the wrong labeling of a QRS complex, and false negative as the missing labeling of an existing complex.
In this section, we show the software re-implementation of the work in \cite{Zanoli:279647}, changing the event-based sampler from level-crossing to our developed PAS. Finally, in Section \ref{sec:system_integration} we present a full system integration of this work, computing the energy performance improvement derived from the integration of the PAS with respect to the uniformly sampled version of the same system.

\subsubsection{Dataset}
The used dataset was extracted from the MIT-BIH Arrhythmia Database \cite{Goldberger00}. This dataset is composed of 48 ECG recordings, each one 30 minutes long, sampled at 360 Hz per channel with an 11-bit resolution over a 10~mV range. Recordings provide 2 ECG channels obtained from 2+1(reference) leads placed on the chest, with the exception of one recording that was recorded using only 1+1 leads. In this work, we use only the signals coming from the MLII lead, since it is the most frequent one in the database. We, therefore, discarded the records where this lead is not present (i.e., record 102 and 104). 

The selected ECG database is considered one of the standard datasets for validation of algorithms operating on ECG signals. This is due to the variability of the signals and the type of heartbeats contained in it, and the complete labeling of all the beats and rhythms by medical experts.

\subsubsection{QRS complex detection}
The main contribution of the work in \cite{Zanoli:279647}  is a novel algorithm for the real-time detection of QRS complexes on event-based sampled signals. Yet, this algorithm has not been designed from scratch, but it is based on one of the most robust methods in the state of the art. The choice was based on two main reasons: the possibility of being re-adapted to a non-uniformly sampled signal, and the ability to design the full stack, from prototyping to implementation (in an emulator). This situation led to exclude all methods involving wavelet or Fourier transformations: even if a theory of wavelet on event-based sampled domain exists, the implementation would require more energy than the adaptation of classic algorithms, and a clear advantage in terms of detection performance has not been proven. 

The considered algorithm is the \texttt{gQRS} algorithm, published in the WFDB software compilation from Physionet~\cite{Goldberger00}, and recognized as one of the best-performing algorithms on public ECG databases~\cite{Llamedo14}.
Other approaches exist for the detection of QRS complexes on event-based sampled signals \cite{Sabzevari2018AnUQ}, but they are fully coupled with the level-crossing sampling strategy. The proposed method works with any non-uniformly sampled signal that can be reconstructed by linear interpolation, and keeps the linear complexity of the original algorithm. These features of the \texttt{gQRS} algorithm allow us to substitute the level-crossing sampler with our PAS without requiring any other major changes. All the details about the implementation of \texttt{gQRS} in the event-based domain are available in~\cite{Zanoli:279647}.

\subsection{Task Recognition from Inertial Signals}
\label{inert}
Almost every smartphone or smartband is nowadays equipped with an Inertial Measurement Unit (IMU). This IMU is composed by, at least, a tri-axial accelerometer and a tri-axial gyroscope. IMUs can be used, without the need for additional devices, for continuous monitoring of human activity \cite{activity_recon_dataset}. This makes inertial signals highly important in modern bio-monitoring applications and a perfect candidate to test our sensing technique. The target task in the proposed experiment is postural and postural transition recognition.

The reference work for this experiment is \cite{Reyes_Ortiz_2016}. We chose this work mainly for three reasons: the good results reported, the non-triviality of the task, and the applicability of the study in a real-world scenario. Next, we briefly analyze the original work, and the key differences with our implementation.

\subsubsection{Dataset}
The used dataset was obtained from \cite{activity_recon_dataset}. This dataset is composed of 30 recordings of subjects with ages between 19 and 48 years. The subjects performed a protocol composed of six basic activities: three static postures (standing, sitting, lying) and three dynamic activities (walking, walking downstairs, and walking upstairs). The experiment also included postural transitions that occurred between the static postures: stand-to-sit, sit-to-stand, sit-to-lie, lie-to-sit, stand-to-lie, and lie-to-stand. All the participants were wearing a smartphone on the waist during the experiment execution. Then, the used signals are composed of the 3-axial linear acceleration and 3-axial angular velocity, acquired at a constant sampling frequency of 50Hz by using the embedded accelerometer and gyroscope of the device. 

The experiments were video-recorded and manually labeled. Each recording is of a different length, varying from three to ten minutes and being six minutes on average. Moreover, the signals are only partially labeled, since no label has been defined for transitions between dynamic activities or between active and dynamic activities.

\subsubsection{Task Recognition Algorithm}
\label{orig inert}
The algorithm analyzed was the one proposed in \cite{Reyes_Ortiz_2016}. In this case, a mobile application was developed to recognize six activities (three static and three dynamic) and six transitions between the static activities, relying exclusively on the IMU of a smartphone. 
The signals were pre-processed by applying noise filters and then sampled in fixed-width sliding windows of 2.56 sec and 50\% overlap (128 samples/window). The acceleration signal, which has gravitational and body motion components, was separated --- using a low-pass filter --- into body acceleration and gravity. The gravitational force is assumed to have only low-frequency components. Therefore a filter with 0.3 Hz cutoff frequency was used. From each window, a vector of 561 features was obtained by calculating variables from the time and frequency domain. This feature set and respective applications can be found in Table \ref{tab:features}.
This leads to 170 to 270 labeled features vector for each signal.

\begin{table}
\begin{center}
\caption{Features and signals used to extract them for the activity recognition task. Time vector refers to all the vectors extracted from the dataset that are defined as a time series, while frequencies vectors are their relative Fourier transform. More details can be found in \cite{Reyes_Ortiz_2016}.}
\label{tab:features}
 \begin{tabular}{|c|c|} 
 \hline
 Feature & Signal applied to \\ 
 \hline\hline
 Mean & Time and frequency vectors \\
 \hline
 Standard deviation & Time and frequency vectors \\
 \hline
 Median absolute deviation & Time and frequency vectors \\
 \hline
 Max & Time and frequency vectors \\
 \hline
 Min & Time and frequency vectors \\
 \hline
 Skewness & Time and frequency vectors \\
 \hline
 Kurtosis & Time and frequency vectors \\
 \hline
 Signal magnitude area & Time and frequency vectors \\ 
 \hline
 Entropy & Time and frequency vectors \\ 
 \hline
 Inter-quartile range. & Time and frequency vectors \\
 \hline
 Max frequency index & Frequency vectors \\ 
 \hline
 Energy & Frequency vectors \\ 
 \hline
 Mean frequency & Frequency vectors \\ 
 \hline
 Energy bands & Frequency vectors \\ 
 \hline
 Auto-regression & Time vectors \\ 
 \hline
 Correlation & Time vectors \\ 
 \hline
 Angle & Time vectors \\ 
 \hline
\end{tabular}
\end{center}
\end{table}

To perform activity recognition, two SVMs were trained for the recognition of only the six main activities without postural transition (six classes) and recognition of the six main activities plus a single class that represent all the postural transition (seven classes).

The train and testing were performed using a slightly modified Leave One Out (LOO) cross-validation method. In classic LOO, given $N$ labeled samples, we train a recognition algorithm to recognize one sample, using as training dataset the other $N-1$ samples. The tested samples are then changed and the whole procedure continues until all $N$ samples are tested \cite{Hastie2009}. In this modified version, we do not train on all samples except one, but instead on all full recordings except a single record. The tested record is then changed, and the procedure continues until all the recordings are tested.

In the original work, the output of the SVM is used as an input to a probability filter to improve the results. In our study, however, we are interested in comparing the performances of the algorithm applied to a uniform signal against an equivalent algorithm operating on an event-based sampled signal. Hence, it is sufficient to compare the results of the SVM --- by choosing the most probable class through an arg-max operation on the output probability vector --- in both scenarios. For this reason, the last stage of the original work was not reproduced.

\subsubsection{Experiment reproduction}
\label{subsec:IMU_reprod}
We redesigned the original work in \cite{Reyes_Ortiz_2016} in two steps. First, we re-sampled the raw signals with our polygonal approximation method. Then, we redesigned the  core-algorithm described in Section~\ref{orig inert} by applying the same signal pre-processing and feature extraction procedure. However, we changed the signal-processing methods from standard, uniformly-sampled-based methods to event-based ones, described in Section~\ref{subsec:methods}. Then, we trained two SVMs with the extracted feature vectors, using the same techniques as the original work. Finally, we compared the results obtained by reproducing the original work analyzed in \cite{Reyes_Ortiz_2016} with our modified signal processing pipeline. These results are analyzed in Section \ref{subsection:IMURes}.

\subsection{Breath Rate Estimation from IR signals}
\label{resp}
The impedance respiratory signal (IR)  describes how the impedance between two or more points on the chest varies in time. This variation is caused by the contraction and relaxation of the chest, measuring the respiration phases of a patient. The objective of this section is to build an event-based algorithm for breath rate detection using IR signals. Thus, we first describe the used dataset. Then, we analyze the custom algorithm we developed for breath rate detection using our event-based approach for wearable sensors.

\subsubsection{Dataset}
The used dataset was obtained from \cite{7748483}. This dataset is composed of several signals, such as photoplethysmography (PPG), impedance respiratory signal, and electrocardiogram (ECG). We are solely interested in impedance respiratory signal, as it is the original signal independently labeled by two experts. Hence, our dataset is composed of 53 signals, each lasting eight minutes and sampled at 125Hz.

Since the IR signals have two sets of labels that can be both considered as ground-truth --- as they were both produced by medical experts ---, the results can be interpreted from both points of view. As a result, we can consider either the first expert to be the ground truth and evaluate the second expert as an optimal (medical level) breath rate detection algorithm, or the opposite. In this work, we treat the second expert as ground truth for the sake of brevity, even if we present the numerical results for both cases.

\subsubsection{Breath Rate Detection Algorithm}
\begin{figure}[h]
  \centering
  \includegraphics[width=0.9\linewidth]{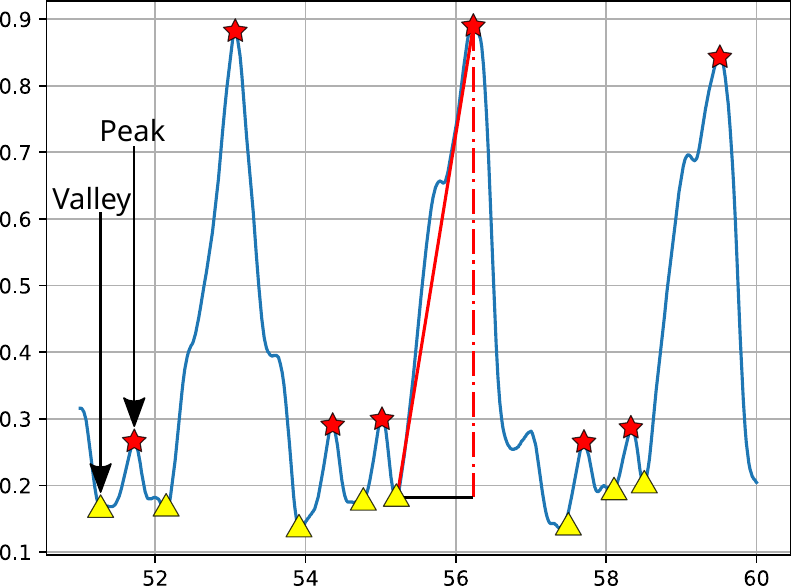}
  \caption{Ten seconds from one of the IR signals. The red stars represent the peaks, while the yellow triangles represent the valleys. Peaks and valleys are used to estimate the slope and delta of a peak (continuous line: slope, dashed line: peak delta)}
  \label{fig:resp}
\end{figure}

Due to the absence of a state-of-the-art algorithm for breath rate detection (using solely the IR signal), we decided to design a simple, but effective, uniform-sampling-based algorithm, presented in Alg.\ref{alg:breath_algo}. An example of the IR signal, as well as the two main features used in the algorithm, can be found in Fig: \ref{fig:resp}.

The main focus of the developed algorithm shown in Alg.~\ref{alg:breath_algo} is to have a simple implementation and results comparable to a human-level labeling. Therefore, we opted for a classic time-domain based, peak detection approach. In particular, the algorithm can be described as follows: for each positive peak structure present in a signal, we search the corresponding valley (negative peak structure) and extract height and slope. We use these two features as the descriptors of the peaks. The list of all peaks is referenced using their descriptors (Alg.~\ref{alg:breath_algo}, line 13). These values are compared with two adaptive thresholds. If the threshold conditions are satisfied, then the peak is inserted into a buffer of candidates (Alg.~\ref{alg:breath_algo}, lines 15-16). After a certain time-window is acquired, we analyze the buffer of candidates and only keep the peaks that meet three constraints (Alg.~\ref{alg:breath_algo}, lines 18-24), namely: 
\begin{itemize}
  \item The peak values (valley-to-peak) need to be higher than half the average of peaks in the buffer.
  \item The slope coefficients need to be higher than half the average of the slope coefficients in the buffer.
  \item The time between peaks need to be longer than the global time threshold.
\end{itemize}
 The thresholds are adjusted by taking 75\% of the previous value and adding 25\% of the average slope and peak of the previous 10 peaks (Alg. \ref{alg:breath_algo}, lines 25,28).

\begin{algorithm}

\footnotesize
\caption{Breath detection algorithm}\label{alg:breath_algo}
\begin{algorithmic}[1]
\Procedure{BreathDetection}{$x(t)$}
    \State $t = 0$
    \State $bufferLen = 10$
    \State $peaks = []$
    \State $peaksBreath = []$
    \State $peakBuffer = []$
    \State $peakCandidates = []$
    \State $globalThresholds = 0$
    \State $localThresholds = 0$
    \State $peaks,globalThresholds$ = $init(x(t))$
    \ForAll{ $p$ in $peaks$}
        \If {$p \in globalThresholds$}
            \State $peakCandidates.add(p)$
        \EndIf
        \If{$Len(peakCandidates)>1 \And t$ is OK}
            \State $localThresholds = computeLocal(peakCandidates)$
            \ForAll{ $pc$ in $peakCandidates$}
                \If{$peakCandidates \in $localThresholds}
                    \State $peaksBreath.add(pc)$
                    \State $peakBuffer.add(pc)$
                \EndIf
            \EndFor 
            \State $peakCandidates.empty()$
            \State $globalThreshold=adjustTheshold(peakBuffer)$
        \EndIf
        \If{$Len(peakBuffer) > bufferLen$}
            \State $bufferLen.remove(first)$
        \EndIf
        \State $t = time(p)$
    \EndFor
\EndProcedure
\end{algorithmic}
\end{algorithm}

After obtaining all the respiration peaks from Alg.~\ref{alg:breath_algo}, we compute an instantaneous breath rate (BR) for every consecutive couple of peaks. We then linearly re-sample the obtained vector of BR every second (i.e., at a frequency of 1 Hz). Finally, we apply a 10-second long median filter, introducing a delay of five seconds, considering the assumption that the BR changes smoothly and slowly over time. 
We then extract the BR vector for the human-level labeling by repeating the previous operations (instantaneous BR extraction, then 1Hz re-sampling, and then 10-second median filter) to the peaks labeled by the experts.

Finally, we ported this algorithm to the event-based domain. In order to do so, we need to execute the same operations. Nonetheless, this time we do it without the assumption of a fixed time between consecutive samples. Such operations are: peak detection, valley detection, slope computing, and time measurement between peaks.
The peak and valley detection operations do not require any information about time but only about the values of consecutive samples, hence, no changes are needed. Slope computing and time measurement both require to compute the time from the information of the event instead of counting the number of samples between the relevant points. Therefore, the adaptation of our proposed Alg.\ref{alg:breath_algo} to work with event-based signals is almost trivial.

\section{Results analysis}
\label{sec:results}

To obtain meaningful results for the above-described experiments, we measure the deviation of an experiment-specific performance merit figure between the original and the event-based experiment. We analyze the $F_1$ score on the QRS complex recognition for the ECG experiment, and the average $F_1$ score among classes for the IMU experiment. In contrast, for the IR experiment, we measure the Root Mean Square Error (RMSE) between the BR vectors computed using Alg. \ref{alg:breath_algo} (for both the uniform and event-based versions) and each of the human-level BR vector.
All the results presented here and in Section \ref{sec:system_integration} can be reproduced from \cite{Experiments}, which is our provided open-source repository for this work.
\subsection{Results on ECG data}
\label{subsection:ECGRes}

We are now interested in evaluating the performance considering the experiment proposed in Section IV.A. The objective of this analysis is to compare the QRS complex detection in EB-sampled signals with state-of-the-art QRS complex detection in uniformly sampled signals.

\begin{figure}[h]
  \centering
  \includegraphics[width=0.9\linewidth]{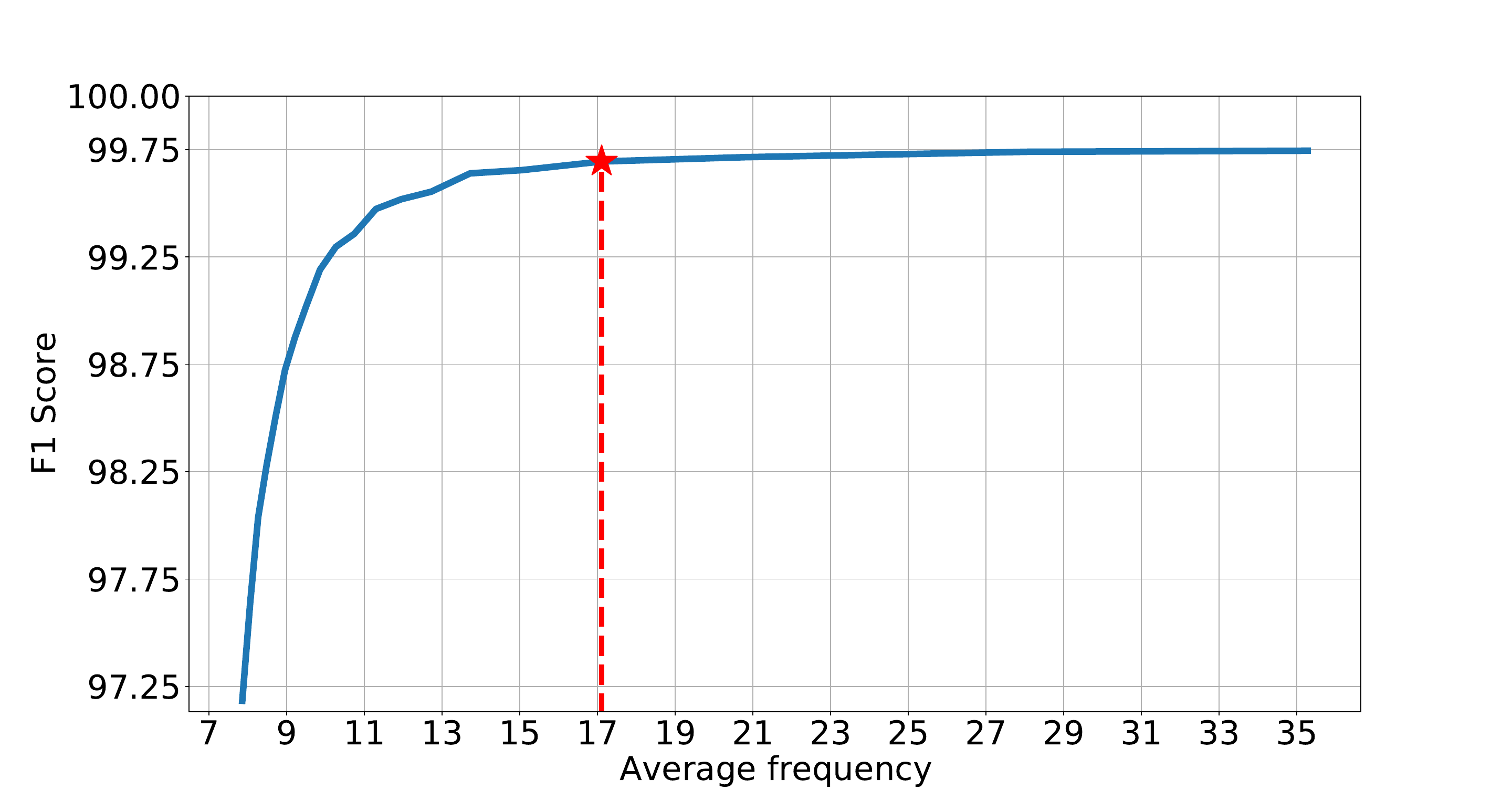}
  \caption{Performance of the event-based QRS detector vs. average frequency. The highlighted point was chosen as the algorithm's working point to have a good compromise between the saved energy and the performance degradation. 
After an average frequency of 35Hz, the methodology reports no improvement when compared to the uniformly-sampled QRS detection algorithm. 
}
  \label{fig:ECG-F1Freq}
\end{figure}

As described in \cite{ECG_quality_assessment}, we cannot restrict the evaluation of ECG quality to a simple binary decision (pass or fail). However, in contrast to the objective in \cite{ECG_quality_assessment}, the main task of our algorithm is an efficient QRS detection, and the signal-to-noise ratio model proposed would not be representative of a correct QRS positioning in the EB signal. 
To measure the performance of QRS detection, we use the ground-truth labels given by the experts in the MIT-BIH Arrhythmia dataset. We match them with the labels found by our algorithm, and then compute the average $F_1$ score of the detection.
The threshold $\epsilon$ used in the PAS has two distinct but related effects in the evaluation metrics. First, as we increase the threshold, the $SRF$ also increases. This is expected and is also shown in the other experiments. The second effect is the reduction of the $F_1$ score. As we can observe from Fig.~\ref{fig:ECG-F1Freq}, there exists an average frequency (directly bounded to the PAS threshold) below which the $F_1$ score greatly decreases while remaining almost constant before. This behaviour is caused by the excessive threshold used in the PAS. As we can observe in Fig. \ref{fig:subTh}, the QRS complex is a clear signature that is preserved by our sampler even when all the other features of the ECG signals get removed by the polygonal approximation. This remains true until the allowed error is comparable to the average area of the QRS complex. Then, increasing the threshold any further results in the loss of the QRS complex signature. Hence, it produces a fast degradation of the measured performance. 

The chosen working point, highlighted in Fig.~\ref{fig:ECG-F1Freq} with a star, shows an $F_1$ score on the QRS complex recognition, for the \texttt{EB-gQRS}, of 99.69\%, compared to the 99.75\% of the original \texttt{gQRS}. The penalty in the $F_1$ score is counter-balanced by an $SRF = 92.7\%$

\subsection{Results on IMU data}
\label{subsection:IMURes}
The objective of this section is to show the impact of event-based sampling and signal processing on inertial data. To achieve this objective, we replicated the study in \cite{Reyes_Ortiz_2016} in both the uniform-sampling and the event-based sampling domain. Then, we measured the performance difference between these two implementations against the obtained $SRF$
As described in Section \ref{subsec:IMU_reprod}, to properly compare the original and the event-based implementation of the work in \cite{Reyes_Ortiz_2016}, we train an SVM to recognize the same seven classes of the original study. Then, we compare the $F_1$ score of the identified tasks against the ground-truth labels given in the dataset, quantifying the task recognition error. To replicate the results, we relied on the same error function described by \cite{Reyes_Ortiz_2016} and reproduced in Eq.~\eqref{eq_error}, where $\alpha_t$ is the arg-max of the SVM output, $g_t$ is the ground-truth, $PT$ is the single class describing all the postural transition, and $UA$ denotes the case where the probability output of all the classes is less than $\frac{1}{N_{classes}}$, and, thus, the activity is not recognized.
\begin{equation}
  \label{eq_error}
  e(t) = \begin{cases}
    0 & \text{if } \begin{cases}
      g_t = \alpha_t ~or\\
      (g_t = PT ~and~ g_{t-1} \neq g_{t+1} ~and \\ (\alpha_t = g_{t-1} ~or~ \alpha_t = g_{t+1})) ~or \\
      (g_t = PT ~and~ \alpha_t = UA)
    \end{cases}\\
    1 & \text{otherwise}
  \end{cases}
\end{equation}
using Eq.~\eqref{eq_error} we can compute the confusion matrix of the leave one out cross-validation and thus obtain the $F_1$ score of the experiment.

\begin{figure}[h]
  \centering
  \includegraphics[width=0.8\linewidth]{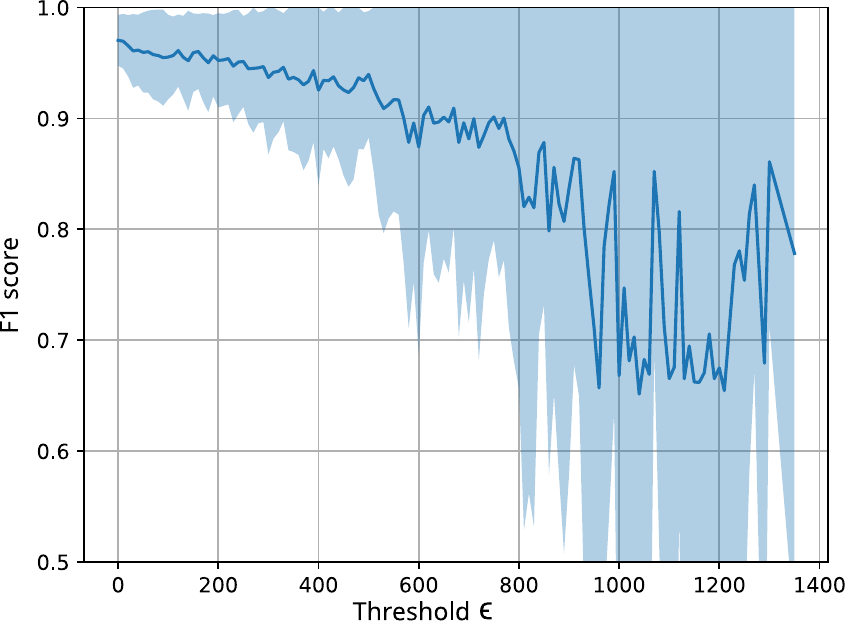}
  \caption{average $F_1$ score and standard deviation of the leave one out cross-validation at increasing thresholding level for the seven classes scenario}
  \label{fig:F1Inert}
\end{figure}

The impact of the event-based sampling on the proposed signal processing pipeline can be evaluated by measuring the average and standard deviation (STD) of the $F_1$ score against the threshold used for the PAS algorithm and comparing these results with the ones obtained using the classic uniform-sampling technique (hence, reproducing the results in \cite{Reyes_Ortiz_2016}). These results are calculated by measuring the $F_1$ score for the recognition of each task. The obtained results are shown in Fig.~\ref{fig:F1Inert}, where we can see that as the threshold increases, the average $F_1$ score decreases, and the STD increases. This is to be expected as a higher threshold means a coarser representation of the signal. Consequently, trivial tasks, composed of big and task-significant signal variations are still well represented and detected, while more complex tasks, composed of smaller and more task-common signal variations are more complex to discriminate. This leads to a lower average and higher STD. 

Also, we observe in this figure that as the threshold increases over a certain point (in Fig.~\ref{fig:F1Inert}, this point is positioned at $\varepsilon=900$), the described behaviour stops. The reason is that at such a high threshold, the PAS algorithm ignores entire chunks of the signal representing tasks that do not induce sufficiently large variations, eliminating such labels from the problem. However, this degradation in performance needs to be correlated to the SRF to understand the true magnitude of what has been observed. The first row of Table~\ref{perf} shows the $F_1$ loss due to the event-based sampling for increasing values of the sampling reduction factor. Correlating these two results, a $SRF=80\%$ results in a threshold of 200 and in a $F_1$ decrease of 3.5\%.

\begin{figure}[h]
  \centering
  \includegraphics[width=0.8\linewidth]{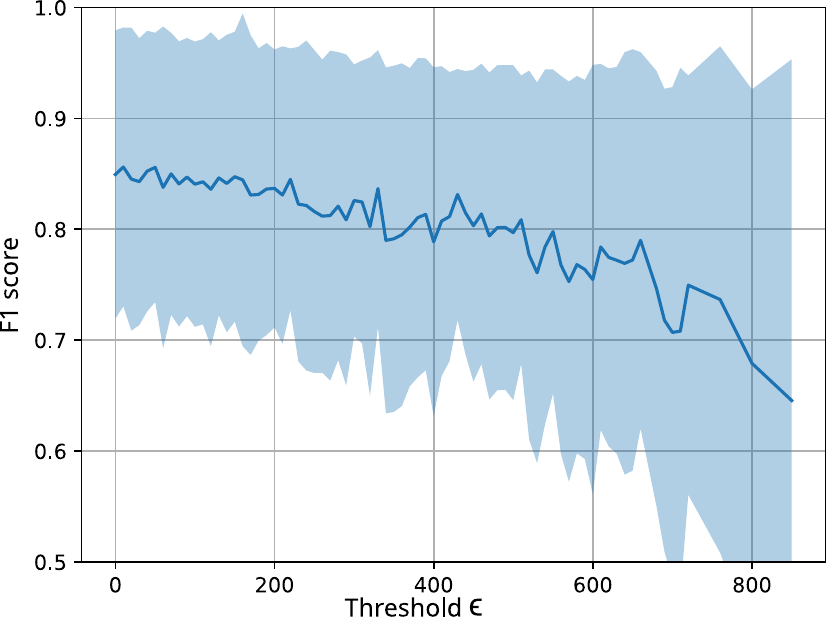}
  \caption{average $F_1$ score and standard deviation of the leave one out cross-validation at increasing thresholding level for the twelve classes scenario}
  \label{fig:F1Inert2}
\end{figure}

Finally, we extended the original study by training a second SVM to recognize all the 12 target classes fully. This extension was performed to explore the potential of the proposed method and extend the original study with interesting results. Moreover, the results obtained from the extended experiments give us a confirmation of the behaviour of porting this type of algorithms (composed of classical signal processing for features extraction) to the event-based domain. 
Minor modifications were needed, as the results for this problem were not computed anymore using Equation~\eqref{eq_error}, but just by taking the arg-max of the probability vector produced by the SVM. Moreover, the SVM had to be modified to process 12 classes instead of 7. 
As for the previous case, we aim at analyzing the impact of the event-based sampling on the signal processing pipeline. Fig.~\ref{fig:F1Inert2} shows how the $F_1$ score varies with the threshold, observing the same behaviour as for the seven classes problem, but starting from a lower baseline value. These results further validate the previously described behaviour. The second row of Table~\ref{perf} shows the $F_1$ loss due to the event-based sampling as the data reduction factor increases.

\begin{table}[h]
\caption{SRF and corresponding $F_1$ loss for the two experiments. The original $F_1$ score for the seven classes problem is 97.1\%, while it is 84.9\% for the 12 classes scenario. A negative loss implies an increase in the $F_1$ score.}
    \resizebox{\columnwidth}{!}{
    \begin{tabular}{l|l|l|l|l}
      $SRF$                            &   20\%   & 40\% & 60\% & 80\% \\ \hline
      seven-classes problem $F_1$ loss &  0.0   & 0.4  & 1.2  & 3.5  
      \\
      12-classes problem $F_1$ loss    &  -0.7   & 0.4  & -0.1  & 1.9               
    \end{tabular}
    }    
    \label{perf}
\end{table}

\subsection{Results on IR data}
\begin{figure}[h]
  \centering
  \includegraphics[width=0.9\linewidth]{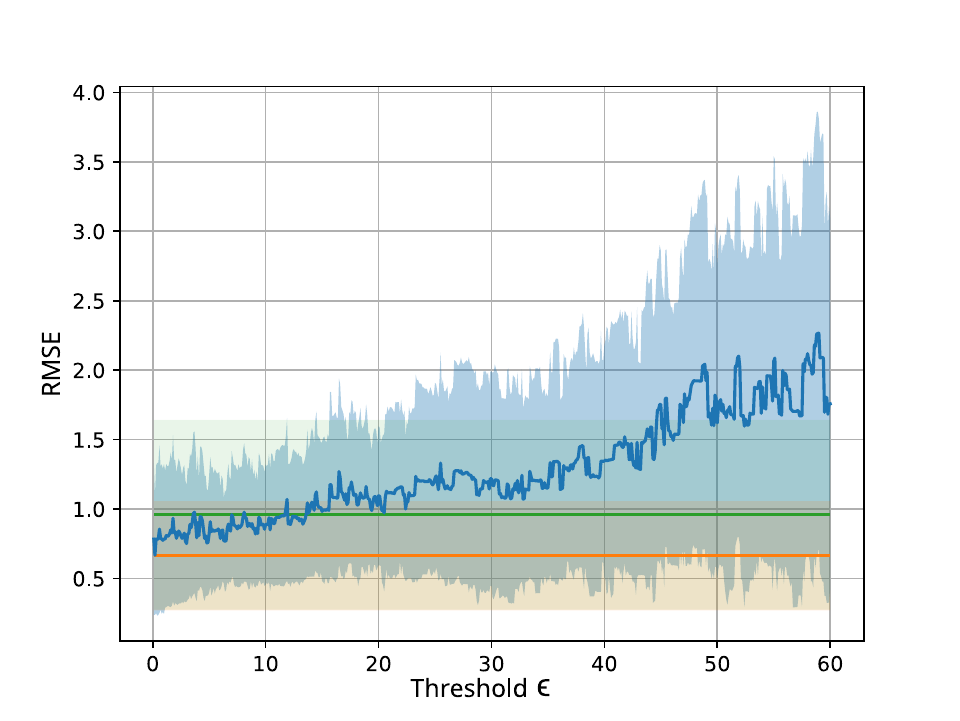}
  \caption{Median and Median Absolute Deviation (MAD). 
 RMS of the detected breath rate, considering the second expert's labels as the ground truth. The orange line corresponds to the first expert, the green to the algorithm on the uniform sampled signal and the blue to the event-based version applied at different threshold levels}
  \label{fig:RMS_second}
\end{figure}

The results obtained for the estimation of breath rate cannot be interpreted as a classification problem, as in the above experiments, as we are interested in evaluating the difference between two vectors with real values. To evaluate the performance of the uniform-sampling-based algorithm and its event-based counterpart, we measure the root mean square error (RMSE) between the estimated BR vector and the  ground truth BR vector. In Fig.~\ref{fig:RMS_second}, we show the performance degradation of the event-based algorithm against the used threshold. As observed in Section \ref{subsection:IMURes}, as the threshold increases the performance degrades and the standard deviation increases. However, by observing the section of Fig.~\ref{fig:RMS_second} below a threshold value of 15, we notice that the average RMS is lower or equal for the event-based version of the algorithm, when compared to the uniform-sampling version. In fact, this signal dramatically benefits from the PAS sampling technique that smooths out peaks and valleys unrelated to the respiration process (mainly caused by artifacts).

In figure Fig.~\ref{fig:RMS_second}, we show a span of thresholds from 0.1 to 60. However, the $SRF$ is already high from the very beginning of this scale: with $\epsilon=0.1$, the $SRF$ is 93.9\%, lowering the average frequency  from 125 Hz to 7.57Hz. If we set $\epsilon=10$, then the $SRF$ results to be 98.9\%, lowering the average frequency of the signal to 1.32Hz, and with $\epsilon=20$, the $SRF$ is 99.2\%, lowering the average frequency to 0.953Hz. These fractional values of average frequencies are to be expected as the regulation is performed on the accepted maximum integral error (the threshold, $\epsilon$). This parameter indirectly affects the number of samples acquired. To directly regulate the average frequency, we should reverse the formulation of the Wall-Danielsson algorithm. This would require fully acquiring the signal until the end, defeating the purpose of this work. This shows that the initial sampling frequency of 125Hz was highly overestimated for this type of signal.

Table~\ref{perf_resp} shows the increase in RMS with respect to both experts for four different $SRF$ values. We observe that choosing the first expert as ground-truth, the average RMSE increase for both the analyzed version. However, the event-based version still shows to perform better than the uniform one. From these results, we can conclude that Alg. \ref{alg:breath_algo} produces results more consistent with the second expert. Also, we can derive that the proposed event-based sampling approach improves the detection of meaningful events in slowly varying signals.

Finally, since the IR dataset used is registered in a medical and controlled environment, the obtained results may not be representative of a day-to-day use of a wearable senor. Hence, we assess the physical noise robustness of the proposed algorithm using the model described and distributed in the MIT-BIH noise stress test database \cite{noise_stress_test}. To model a noisy IR signal we first weigh the three noise source provided by this database as in eq~\ref{eq:noise} (with $Noise_{mio}$ being the mioelectric muscular noise, $Noise_{mov}$ the noise caused by movement and electrode misplacing, and $Noise_{base}$ the baseline wandering noise, already present and filtered in the non-noisy original signal ) to obtain a model representative of the noise registered by chest mounted impedance sensor. The obtained noise is scaled to obtain the desired value of signal to noise ratio (SNR) compared to the database IR signal. Contrary to what proposed in the noise stress test database web page, we compute the noise scale factor measuring the complete signal and noise average power instead of the five-minute window measurements proposed. We then invert the SNR definition formula to obtain the noise scaling factor and apply the obtained noise to the whole signal. This corresponds to the worst possible scenario for the designated SNR value:
\begin{equation}
    \label{eq:noise}
    noise = 2*Noise_{mio}+5*Noise_{mov}+1*Noise_{base}
\end{equation}

\begin{figure}[h]
  \centering
  \includegraphics[width=0.9\linewidth]{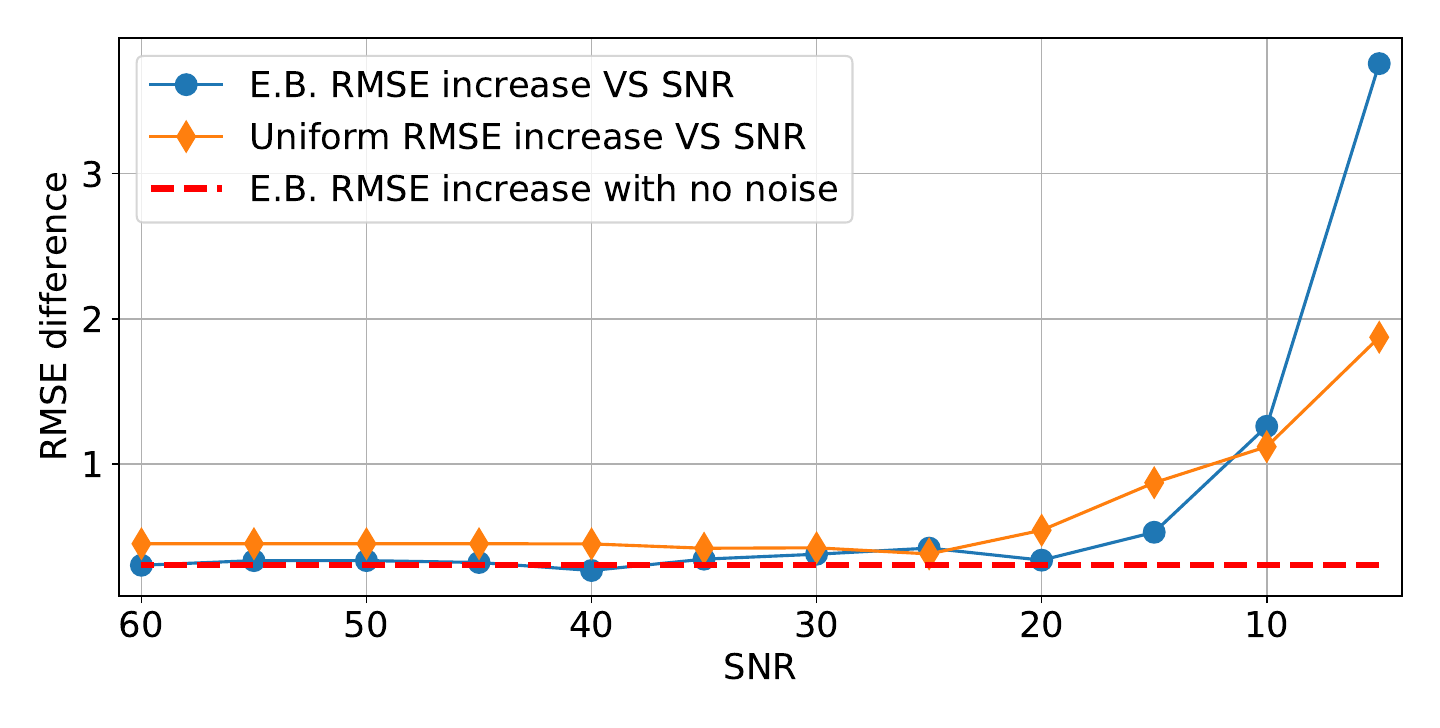}
  \caption{RMSE increases with respect to the reference human RMSE with increasing noise (decreasing SNR). Dashed line: EB algorithm without additional noise. Circles: RMSE increment for the EB version of the algorithm with $\varepsilon=2$. Diamonds: RMSE increment for the uniformly sampled version.}
  \label{fig:RMS_SNR}
\end{figure}

As we can observe in figure \ref{fig:RMS_SNR}, the algorithm results to be robust (i.e., without significant effect on the results) up to an SNR of 20 dB. Moreover, the EB-version of the algorithm continues to outperform the uniform version until an SNR of 10 dB. While both algorithms start to degrade with an SNR lower than 20 dB, the intrinsic filtering of the PAS algorithm helps to remove high-frequency noise, increasing the robustness of the EB-algorithm below 10~dB of SNR.

\begin{table}[h]
\caption{Sampling reduction factor (SRF) and corresponding RMS increase for the two scenarios (expert one or expert two considered as ground truth). The median RMS increase for the original algorithm on the uniform-sampled signal is of 0.5 with respect to the first expert, and 0.3 with respect to the second expert.}
    \resizebox{\columnwidth}{!}{
    \begin{tabular}{l|l|l|l|l}
      $SRF$                       & 96\% & 97\% & 98\% & 99\% \\ \hline
      Expert 1 as ground truth    & 0.26 & 0.30 & 0.36 & 0.49  
      \\
      Expert 2 as ground truth    & 0.12 & 0.13 & 0.17 & 0.29               
    \end{tabular}
    }    
    \label{perf_resp}
\end{table}

\section{System Integration}
\label{sec:system_integration}

In the previous sections, we have first described the circuital implementation of the PAS algorithm and then shown the applicability and potential benefits of event-based sampling in signal processing to different scenarios. In this section, we propose a full system simulation applied to the first targeted problem, described in Section \ref{subsec:QRS_ECG}: on-line, low-power QRS complex detection from ECG signals. We discussed the performance results in Section \ref{subsection:ECGRes}. In this section, we summarize the baseline implementation from which we departed in our previous work \cite{Zanoli:279647}. Then, we present the new experimental setup, the improvements in the designed system, and the energy results obtained.
 
\subsection{Baseline implementation}
\label{subsec:prev_work}
In order to measure the energy consumed and the performance achieved, both the implemented algorithms (\texttt{gQRS} and \texttt{EB-gQRS}) have been implemented on a low-power, high-performance microcontroller unit (MCU) based on the PULP platform. This MCU, called Mr.Wolf \cite{Pullini18}, is composed of a RISC-V single-core controller and an octa-core \mbox{RISC-Y} cluster. This platform has been developed using $40~nm$ technology. Both algorithms, first, acquire a 20-second signal window through a \textmu DMA. Then, the algorithm is executed with this window as the input signal. Finally, the MCU is again put in sleep mode, waiting for the next window.

The results of this previous work can be divided according to two main objectives:
\begin{itemize}
    \item \textbf{Detection performance}: As presented in Section \ref{subsection:ECGRes}, the \texttt{EB-gQRS} achieves an $F_1$ score on the QRS complex recognition of 99.69\%, compared to the 99.75\% of the original \texttt{gQRS}. The worsened $F_1$ score is counter-balanced by an $SRF = 92.7\%$
    \item \textbf{Energy consumption}: The baseline implementation of the \texttt{EB-gQRS} greatly reduces the energy for the execution of the QRS complex detection algorithm. 
    At the same time, the energy used to keep the MCU in sleep mode, while the acquisition process is completed, remains invariant between the two algorithms (EB-gQRS, and gQRS) and dominates the total energy consumption, resulting in a total energy reduction of 44.6\%. Moreover, the energy computation obtained here does not take into account the energy requirements for power-state switching. To the best of our knowledge, the energy required in-between power modes is not reported in~\cite{MrWolf}. In view of the overall energy requirement of the system, we consider this contribution to be marginal.

\end{itemize}

\subsection{Experimental setup}
Starting from the aforementioned work, we replace the level-crossing sampler by the proposed PAS. Then, we extend the energy results obtained with a more detailed analysis of the whole system. In this case, we also consider the energy consumed by the sampler, instead of analyzing only the energy used for saving sampled data and processing. 

Finally, we used the platform described in Section \ref{subsec:prev_work}: Mr.Wolf \cite{Pullini18}, but modifying its memory structure, moving from 64~KB to 2~KB memory blocks. This reduction in the memory block sizes was made because we observed the largest impact on energy consumption was due to the leakage of the memory blocks when they are in a retentive state. By modifying the memory block size, we can reduce this impact as the leakage of a smaller memory block is, in a first approximation, directly proportional to the reduction factor. Moreover, even in the original implementation, the amount of data stored never grew more than 3~KB. By applying this change, we switch from one 64~KB block of memory to two 2~KB blocks. Here, we refer to the acquired data without considering the size of the program in memory, although the program memory footprint is considered for the energy consumption computation.

The power analysis of our designed PAS was run on Synopsys Design Compiler, using the design synthesized on a $40~nm$ technology library. We followed the same experimental procedure as \cite{Zanoli:279647}. First, we used the PAS algorithm to event-based sample the MIT-BIH Arrhythmia Database at increasing threshold levels. Then, we measured the $F_1$ score and energy consumption of the proposed algorithm for each threshold. Finally, we measured the same parameters for the original \texttt{gQRS} algorithm working on the uniformly sampled signal and compared the results.
As in our previous work, the algorithm is run on non-overlapping windows of 20 seconds.

\subsection{Experimental energy and performance exploration results}

We evaluated our design using technology libraries that correspond to the worst operating case. In the context of an embedded wearable device, the activity is a sequence of short computation and long sleep periods. We thereby expect the energy consumption to be dominated by leakage. Consequently, we consider the technology corner that maximizes the leakage (Fast-Fast corner) and we kept the temperature in a reasonable range for a wearable ($25~\degree C$).

Table \ref{tab:synthesis_info} shows the results indicating that the PAS has an active power consumption of $1.24~\mu W$ and a leakage power of $0.428~\mu W$. Due to the speed at which the PAS processes every input sample ($7.09~ns$ per sample), the active energy, on a 20-second window, is negligible when compared to the leaked one: $63.1~pJ$ against $8.56~\mu J$. This translates to a fixed PAS consumption of $8.56~\mu J$ per acquired window.

The \texttt{EB-gQRS} algorithm consumes a variable amount of energy, depending on the threshold (and hence on the average frequency) used in the PAS. Indeed, it varies from $143~\mu J$ at 10~Hz to $379~\mu J$ at 360~Hz. 

\begin{figure}[tp]
	\centering
    \includegraphics[width=1\linewidth]{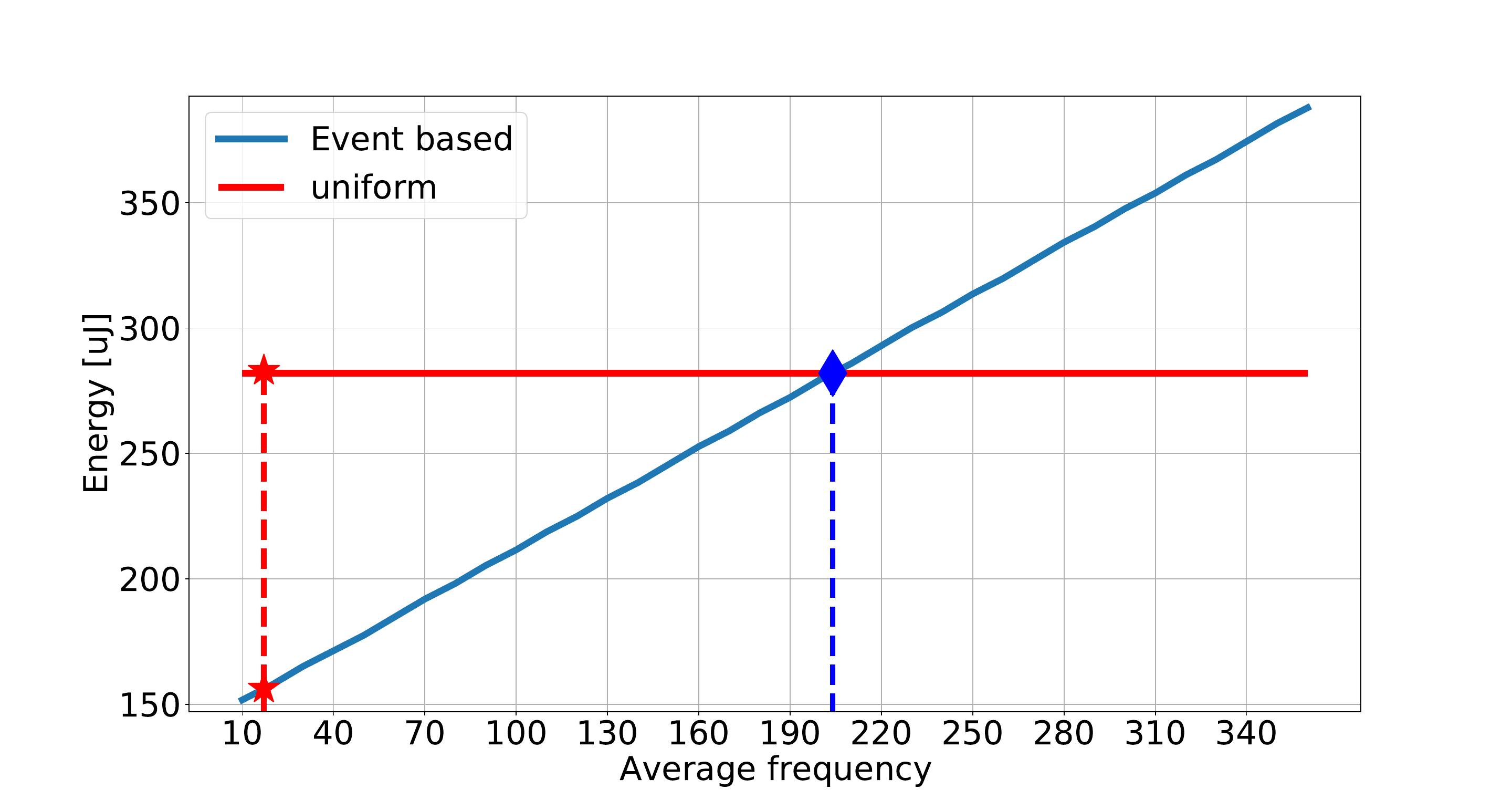}
	\caption{Energy comparison between the original \texttt{gQRS} algorithm and the event-based version. The red star represents the working point, 17.1~Hz, the blue diamond is the intersection point at 204 Hz.}
	\label{fig:energyCompareson}
\end{figure}

Fig. \ref{fig:energyCompareson} shows the energy consumption comparison between the \texttt{gQRS} algorithm and the event-based version plus the energy used by the PAS, at different average frequencies. While the average frequency of the PAS algorithm remains below 204~Hz, the event-based algorithm uses less energy than the original one. Moreover, in the range between 10 and 30 Hz, the energy consumption measured is reduced by 40\%.

\begin{figure}[tp]
	\centering
    \includegraphics[width=1\linewidth]{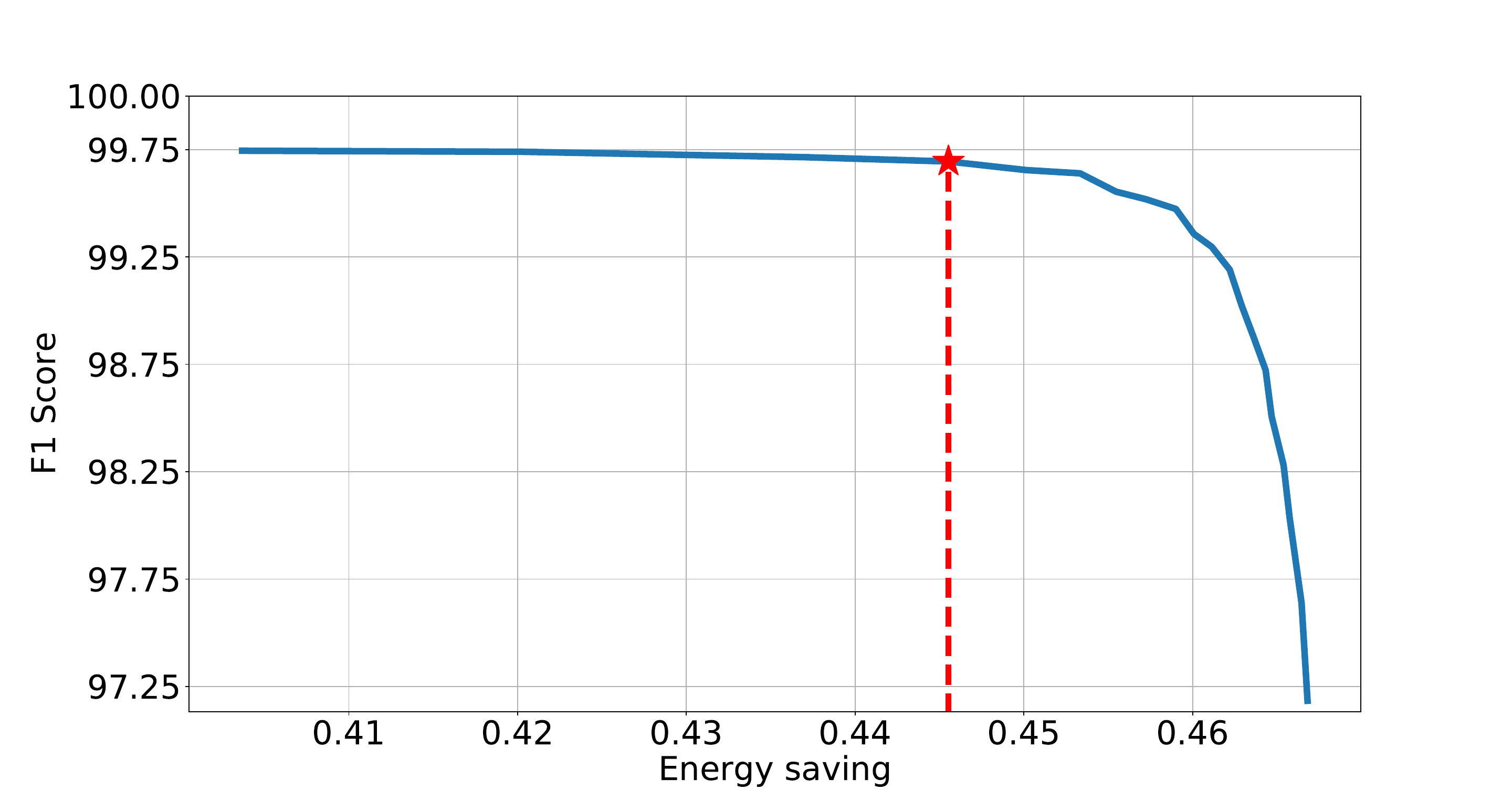}
	\caption{Performance of the event-based QRS detector vs. energy savings. The highlighted point was chosen as a working-point for the algorithm to obtain a good compromise between the saved energy and performance degradation. Since the energy reduction is a direct consequence of the average frequency reduction, we can notice how this figure presents a mirror view to Fig.~\ref{fig:ECG-F1Freq}.
}
	\label{fig:scoreVSenergy}
\end{figure}

In a complementary view, Fig.~\ref{fig:scoreVSenergy} shows the degradation in the QRS detection performance, in terms of $F_1$ score, as the energy savings increase. In this sense, while the original \texttt{gQRS} algorithm achieves an $F_1$ score of 99.75\%, the highlighted point --- corresponding to an average frequency of 17.1 Hz ---  shows energy savings of 44.6\%, while achieving an $F_1$ score of 99.69\%. When compared to the original sampling frequency of 360~Hz, these results translate in a worsening in $F_1$ score of 0.06\% and $SRF = 95.2\%$

The presented energy consumption values for the PAS are computed assuming maximum leakage, Regular Voltage Threshold (RVT), and considering the working temperature to be $25~\degree C$. 
In these conditions, this energy is already small relative to the consumption of the used MCU in the selected working point. However, it is possible to lower even more the consumption of the proposed sampler aiming specifically at reducing the energy leaked by the flip-flops while in the retentive state. By changing the technology from RVT to High Voltage Threshold (HVT) we managed to reduce the leaked power from $0.428~\mu W$ to $0.035~\mu W$. This would bring the total energy needed by the PAS for acquiring the 20-second windows of data to $0.7~\mu J$, achieving  energy savings of 47.3\% at the designed working point.

\section{Conclusion}

Interconnected smart sensors are paving the way of Industry 4.0, and the combination with low-power wearable technology can enable a much healthier and safer working environment. However, a significant energy of the system is spent in standard sensing approaches, which is not suitable for a sustainable IoT world. In this paper, we have analyzed, designed, and validated our Polygonal Approximation Sampler (PAS), which is a novel circuit for event-based sensing relying on a polygonal approximation method targeting low-power wearable sensors. The design was done through logic synthesis and the energy evaluation was performed using a TSMC $40~nm$ technology library, which resulted in an active power consumption of $1.24~\mu W$ and a leakage power of $0.428~\mu W$. The circuit has also been field-tested by targeting three common signal processing tasks on three different types of archetypal signals captured by wearable devices (i.e., accelerometers, respiration data, and ECG). Moreover, we have compared our PAS design with standard periodic ADC sampling. Overall, we obtained a reduction in the number of acquired samples from 60\% to 99\% without significant performance penalties. Finally, we evaluated the energy consumption of a system integrating the PAS with a complete SoC platform, showing a total reduction in energy consumption of 44.6\%. These results show that the PAS is a promising component to be integrated into a wide range of wearable devices performing continuous acquisition and analysis of different types of sensory signals. Thus, PAS paves the way for a better and safer interaction between workers and machines in the future of Industry 4.0.




\section*{Acknowledgment}

This work has been supported in part by Swiss NSF ML-Edge Project (GA No. 182009), the  EC H2020 DeepHealth Project (GA No. 825111), and the EC H2020 DIGIPREDICT Project (GA No. 101017915).

\ifCLASSOPTIONcaptionsoff
  \newpage
\fi



%
\bibliography{references}
\bibliographystyle{ieeetr}

%








\end{document}